\documentclass[pre,showpacs,twocolumn,amsmath,amssymb,superscriptaddress]{revtex4-1}

\usepackage{graphicx,color}
\usepackage{amsmath,amssymb,latexsym}
\usepackage{pgfplots}
\usepackage{mathtools}

\newcommand{\mean}[1]{{\left< #1 \right>}}
\newcommand{\diff}{{\rm d}}

\begin{document}


\title{Energy repartition for a harmonic chain with local reservoirs}

\author{Gianmaria Falasco}
\email{falasco@itp.uni-leipzig.de}
\affiliation{
Institut f\"ur Theoretische Physik, Universit\"at Leipzig,  Postfach 100 920, D-04009 Leipzig, Germany}

\author{Marco Baiesi}
\affiliation{
Department of Physics and Astronomy, University of Padova, Via Marzolo 8, I-35131 Padova, Italy
}
\affiliation{
INFN, Sezione di Padova, Via Marzolo 8, I-35131 Padova, Italy
}

\author{Leo Molinaro}
\affiliation{
Department of Physics and Astronomy, University of Padova, Via Marzolo 8, I-35131 Padova, Italy
}

\author{Livia Conti}
\affiliation{
INFN, Sezione di Padova, Via Marzolo 8, I-35131 Padova, Italy
}

\author{Fulvio Baldovin}
\email{baldovin@pd.infn.it}
\affiliation{
Department of Physics and Astronomy, University of Padova, Via Marzolo 8, I-35131 Padova, Italy
}
\affiliation{
INFN, Sezione di Padova, Via Marzolo 8, I-35131 Padova, Italy
}
\affiliation{
Sezione CNISM Universit\`a di Padova,
 Via Marzolo 8, I-35131 Padova, Italy
}

\date{\today}

\begin{abstract}
We exactly analyze the vibrational properties of a chain of harmonic
oscillators in contact with local Langevin heat baths. Nonequilibrium
steady-state fluctuations are found to be described by a
set of mode-temperatures, independent of the strengths of both the harmonic
interaction and the viscous damping. 
Energy is equally distributed between
the conjugate variables of a given mode but differently among
different modes, in a manner which depends exclusively on the bath
temperatures and on the boundary conditions. 
We outline how
bath-temperature profiles can be designed to enhance or
reduce fluctuations at specific frequencies in the power spectrum
of the chain length.

\end{abstract}

\maketitle

\section{Introduction}

The enhancement of nonequilibrium fluctuations at low wavenumbers is a
key feature of systems driven by thermodynamic gradients (see
\cite{deZarate_2006} for a review). For temperature gradients, it has
been thoroughly studied both theoretically \cite{Schmitz.1988}, and
experimentally in systems ranging from simple fluids
\cite{Sengers_PRA1992} to polymer solutions \cite{Sengers_PRL1998} and
fluid layers also under the influence of gravity
\cite{Takacs_PRL2011}. More recently, fluctuations in nonisothermal
solids have been the subject of experimental investigation, fostered
by the possibility of technological applications in fields as diverse
as microcantilever-based sensors \cite{Harris.2008} and gravitational wave detectors
\cite{conti}. For example, the low frequency vibrations of a metal
bar, whose ends are set at different temperatures, were found to be
larger than those predicted by the equipartition theorem at the local
temperature \cite{conti_01}, thus corroborating the generality of the
results obtained for nonequilibrium fluids \cite{monotonic_dispersion}.
(See also experiments with cantilevers~\cite{bellon}).
 
Theoretical studies of nonequilibrium solids
focused more on 
thermal conduction in low dimensions, where crystals
are usually modeled as Fermi-Pasta-Ulam oscillator chains coupled at
the boundaries with heat baths at different temperatures
\cite{lepri_97,lepri_03,dhar_08}. Thanks to their simplicity, 
integrable and quasi-integrable
models may be taken as a paradigm to describe more comprehensively the
energetics of normal solids under nonisothermal conditions. 
For instance, anomalous features are known to disappear when,
in place of non-homogeneous boundary
conditions at the borders, a
temperature gradient is generated by stochastic heat baths displaced
along the system~\cite{dhar_08}. 
Specifically, it has been shown that self-consistent heat baths -- such 
that no energy flows on average into or out of the
reservoirs -- are sufficient to recover the Fourier's law of heat conduction
in a harmonic chain \cite{bolsterli_01,bonetto_01,dhar_06}. Lifting
the ``self-consistency" condition, one obtains a simple, yet general,
model which describes a solid immersed in a locally
equilibrated medium~\cite{falcao_06,fogedby_14,imparato_15}. 
This can find application in all cases where the
study of fluctuations is applied to an extended system with a complex
thermal balance. 
As an example, we may cite cryogenic gravitational wave
detectors, where thermal fluctuations of the systems
composed by the test masses and their multistage suspension chains
are of central importance. 
The latter are effectively coupled to different heat baths and flows
\cite{Kagra}.

\begin{figure}[!b]
\includegraphics[width=.92\columnwidth]{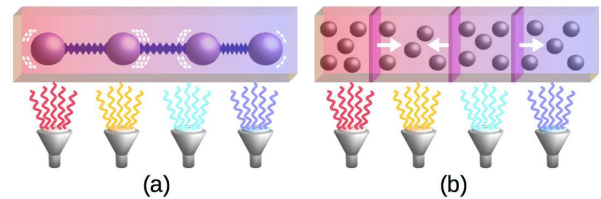}
\caption{(a) Sketch of the linear chain of $N$ harmonic oscillators
  held in a temperature gradient: Each oscillator is
  coupled to an independent heat bath at temperature $T_n$, $n = 0,
  ..., N$ (in the picture $N$~=3). (b) Schematic interpretation in
  terms of sound propagation in a medium.}
\label{fig_sketch}
\end{figure}

Here we analyze the energy repartition among the elastic modes of a
harmonic chain held in temperature gradient, as sketched
in Fig.~\ref{fig_sketch}(a). In a coarse-grained picture, the
oscillator displacements can be thought of as the local strain of a
(one-dimensional) elastic dispersive body, such that the model
describes the damped propagation of thermal phonons
(Fig.~\ref{fig_sketch}(b)).  Our approach is fully analytic and
provides an explicit expression for the energy repartition among the
modes in terms of their effective temperatures 
$\mathcal{T}_{k k}$. 
Exemplifying our results for temperature profiles with a
defined concavity, we show that $\mathcal{T}_{k k}$'s depend only
on this concavity and on the boundary conditions of the
system. 
A naive expectation could be that deviations from energy equipartition
are to be anticipated at long wavelengths only, since local equilibrium conditions
should hold at short scales. 
On the contrary, we find that both long and short wavelength modes can
either heat up or cool down well beyond the average temperature.  
We also study a reverse-engineering approach
in which the heat bath temperatures are inferred starting from a desired
energy repartition.

\section{Model and general results}

Consider a linear chain of $N+1$ equal oscillators located at
positions $q_n$ ($n=0,1,\ldots,N$).  Successive masses are connected
through a harmonic potential of equilibrium length $l_0$. Each of them
is in contact with a specific Langevin bath at temperature $T_n$
\cite{continuous}, providing viscous damping with coefficient $\gamma$
and thermal noise $\xi_n$.  Setting masses to unity, the equations of
motion in the displacement coordinate $R_n\equiv q_n-n\,l_0$ read
\begin{align}
\label{LangevinR}
\ddot R_n 
&=  -\gamma \dot  R_n -\kappa \sum_{m=0}^N A_{n m} R_m + \xi_n,
\end{align}
where $A_{n m}$ is a tridiagonal matrix accounting for first-neighbors
interactions via the potential $\frac \kappa 2 (R_m-R_{m-1})^2$.  In
Eq.~\eqref{LangevinR} the standard Gaussian white noise $\xi_n$ has an
amplitude given by the fluctuation-dissipation theorem at the local
temperature (in units of $k_B$):
\begin{equation}
\langle \xi_m(t)\,\xi_{n}(t')\rangle= 2\gamma\,T_n\;\delta(t-t')\;\delta_{m n}.
\end{equation}

In the following, we first consider the case of free boundary
conditions ($A_{00}=A_{N N}=1$); fixed ($A_{00}=A_{N N}=0$) and mixed
($A_{00}=1$, $A_{N N}=0$) boundary conditions are discussed later in
Appendix~\ref{app:A}.  With free boundaries the matrix $A_{n m}$ is
diagonalized by the linear transformation $\Phi^{-1}A\Phi$, with
\begin{equation}
\label{Phi}
  \Phi^{-1}_{kn}= \frac{1}{N+1} \cos\left( \frac{k\pi}{N+1} \Big(n+\frac1 2\Big)\right),
\end{equation}
mapping the spatial coordinates $R_n$ into the coordinates of
the normal modes $X_k \equiv \sum_n \Phi_{kn}^{-1} R_n$, for which
\begin{equation}\label{LangevinX}
\ddot X_k= -\gamma\,\dot X_k -\omega_k^2\, X_k + \eta_k,
\end{equation}
where $\omega_k^2 = 4\kappa\,\sin^2\left(\frac{k \pi}{2(N+1)}\right)$ 
is the (squared) eigenfrequency of the $k$-th mode.
In this dynamics, the only source of correlation between modes is contained in 
the transformed Gaussian white noises
$\eta_k\equiv \sum_n \Phi_{kn}^{-1}\,\xi_n$,
\begin{equation}
\langle \eta_k(t)\,\eta_{k'}(t')\rangle=2\gamma\,\mathcal{T}_{k k'}\,\delta(t-t')/(N+1),
\end{equation}
These correlation include a ``temperature'' matrix 
\begin{equation}\label{Teff}
\mathcal{T}_{k k'}\equiv
(N+1)
\sum_{n=0}^N
\Phi^{-1}_{kn}\,\Phi^{-1}_{k'n}\,T_n\,,
\end{equation}
which is certainly diagonal only in the equilibrium case
\mbox{$T_n=T\;\;\forall n$}, where energy equipartition is recovered.
In a nonequilibrium state, generated by heterogeneous 
bath-temperatures, the diagonal $\mathcal{T}_{k k}$ still encodes 
information
about how energy is distributed among the modes.
Non-zero off-diagonal
$\mathcal{T}_{k k'}$ emerge in connection with energy fluxes.
To show this, we consider the average kinetic energy ($K_k$) and potential energy ($V_k$) of the $k$-th mode,
\begin{align}
& \  K_k\equiv(N+1)\,\langle \dot{X}_k^2\rangle\,(1-\delta_{k0}/2) ,
&& V_k\equiv(N+1)\,\omega_k^2\,\langle X_k^2\rangle, \nonumber
\end{align}
where expectation values
$\langle\cdot\rangle$ are taken over different realizations of the
thermal noise $\xi_n$.
We get the variances $\langle X_k^2\rangle$, $\langle \dot X_k^2\rangle$
from the solution of Eq.~\eqref{LangevinX} (Appendix~\ref{app:B}), 
\begin{align}
X_k(t)&= \sum_{\alpha=1,2}\int_{-\infty}^{t} d t'
\frac{(-1)^{\alpha}}{\lambda_k^1-\lambda_k^2} e^{-\lambda_k^\alpha(t-t')}\eta_k(t'), 
\end{align}
where $\lambda^{(\alpha)}_k$ with $\alpha=1,2$ are the roots 
of the characteristic equation for the unforced harmonic oscillator;
namely, 
$\lambda^{(\alpha)}_k = -\frac{1}{2} [\gamma +(-1)^{\alpha} \sqrt{\gamma^2 -4\,\omega_k^2}]$.
For each mode $k$, both the average
kinetic and potential energy turn out to coincide with one half of the mode temperature
(Appendix~\ref{app:B}):
\begin{equation}
\label{eq_en_rip}
K_k^X=V_k^X={\cal T}_{k k}/2\quad(k\neq0).
\end{equation}
This relation establishes a form of energy equipartition between the conjugate variables
of a single mode.
Interestingly, from \eqref{Teff} and \eqref{Phi} one sees that 
${\cal T}_{k k}$ does not depend on the details of both
the harmonic interaction ($\kappa$) and the damping ($\gamma$). Therefore, the 
amount of energy stored in the $k$-th mode is directly determined by the choice
of the bath temperature profile $T_n$, for given boundary conditions.
Put in other words, properly designing thermal profiles it is in principle possible to 
enhance or reduce the thermal vibrations
of specific modes.
All these findings are confirmed by numerical integration of~\eqref{LangevinR}.

\begin{figure}
\includegraphics[width=.95\columnwidth]{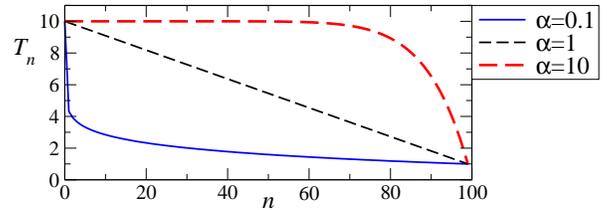}\\
\caption{
  Heat-bath profiles utilized for exemplifying our results.  
}
\label{fig_profile}
\end{figure}

\section{Role of the boundary conditions}

\begin{figure}
\includegraphics[width=.99\columnwidth]{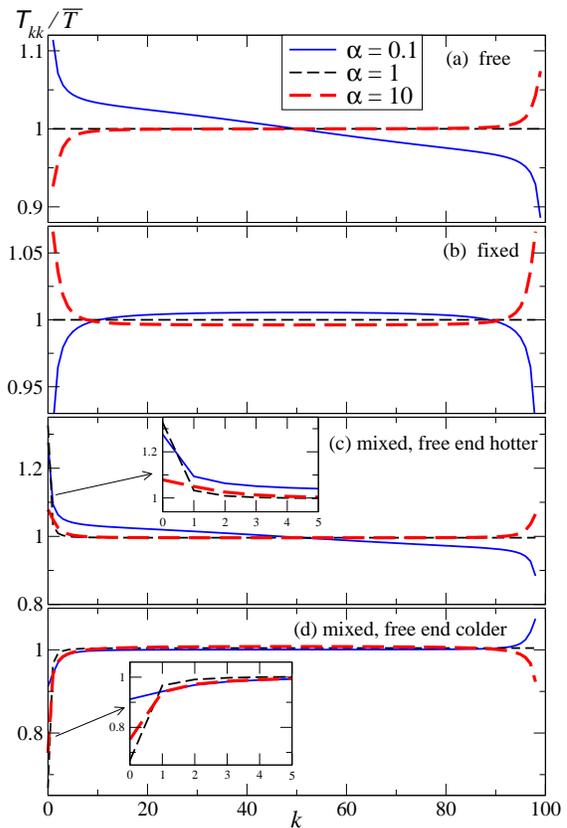}\\
\caption{ Modes normalized temperatures ${\cal T}_{k k}/\overline T$
  calculated through Eqs.~\eqref{Tkk}, \eqref{Tkk_fix},
  \eqref{Tkk_mix} for the heat-bath profiles mentioned earlier.  
  The four panels refer to different boundary conditions:
  (a) free, (b) fixed, (c) mixed with free end hotter than the fixed
  end, and (d) vice versa. Insets enlarge the plots at low $k$'s.  }
\label{fig_4cases}
\end{figure}

In the case of free boundaries, Eq.~\eqref{Teff} gives
\begin{equation}\label{Tkk}
  {\cal T}_{k k}=
{\overline T}
\,\left[
1+ \frac{\sum_{n=0}^N T_n
\cos\left(\frac{2n+1}{N+1}\,k\pi \right)}{\sum_{m=0}^N T_m}
\right]
\;(k\neq0),
\end{equation}
where ${\overline T\equiv\sum_{n=0}^N T_n/(N+1)}$
is the average imposed temperature.
The center-of-mass kinetic energy ${(N+1)\langle\dot{X_0}^2\rangle/2}$
is equal to ${\cal T}_{00}=\overline T$.
Notice that Eq.~\eqref{Tkk} is valid 
in particular when $T_n$ corresponds to a self-consistent profile 
\cite{bolsterli_01,bonetto_01,dhar_06}. 
In~\eqref{Tkk} the energy stored by the mode $k$ under
stationary nonequilibrium conditions emerges like a correction to the
average temperature $\overline T$, which at most amounts to $\pm \overline T$.
This correction
can be viewed as a weighted average of a cosine function
over the temperature profile: For parity, it vanishes for 
all temperature profiles which are odd with respect to $(N/2,\overline T)$.
The relevant physical consequence is that with free boundary conditions energy
equipartition is extended to all nonequilibrium temperature profiles which are
odd-symmetric with respect to $(N/2,\overline T)$, like linear profiles. 
At variance, if the temperature profile has a definite upwards
(downwards) concavity in the interval $[0,N]$, low- (high-) $k$ modes
heat up and high- (low-) $k$ modes freeze down.
We exemplify these findings assuming heat-bath temperatures
$T_n=T_0+(n/N)^\alpha\,(T_N-T_0)$ with $T_0=10$, $T_N=1$, and $N=99$
(see Fig.~\ref{fig_profile}): $\alpha=1$ corresponds to a linear
temperature profile, whereas $\alpha<1$ ($\alpha>1)$ corresponds to a
profile with upwards (downwards) concavity. In Fig.~\ref{fig_4cases}(a)
one finds the resulting $T_{k k}$ for free boundary conditions.

Transport properties might depend crucially on the boundary conditions~\cite{kundu_10}.
We show that the latter strongly influences also the repartition of energy among the normal modes. 
For fixed boundary conditions, ${\cal T}_{k k}$ becomes (Appendix~\ref{app:A})
\begin{equation}\label{Tkk_fix}
  {\cal T}_{k k}=
\frac{
(N-1)\,
{\overline T}
}{
N}
\,\left[
1- \frac{\sum_{n=1}^{N-1} T_n
\cos\left(\frac{2n\,k}{N}\,\pi \right)}{\sum_{m=1}^{N-1} T_m}
\right]
\end{equation}
($0<k<N$), with $\overline T\equiv\sum_{n=1}^{N-1} T_n/(N-1)$.
Fig.~\ref{fig_4cases}(b) shows the mode energy repartition for the
same profiles $T_n$ used for open boundary conditions in
Fig.~\ref{fig_4cases}(a). 
Notably, the low-$k$ behavior is inverted. 
For instance, while free boundaries 
enhance the long-wavelength energy storage for concave-up $T_n$, 
fixed boundaries do the opposite. Hence, if the aim were to store
energy at low $k$'s, a convenient strategy 
would be to heat up the boundaries and cool down the middle 
of a free chain, and vice versa with a fixed chain.

For mixed boundaries in which we leave free the mass at $n=0$ 
and fix the mass at $n=N$ we have (Appendix~\ref{app:A})
\begin{equation}\label{Tkk_mix}
  {\cal T}_{k k}=
\frac{
2N\,{\overline T}
}{
2N+1
}
\,\left[
1+ \frac{\sum_{n=0}^{N-1} T_n
\cos\left(\frac{(2n+1)\,(2k+1)}{2N+1}\,\pi \right)}{\sum_{m=0}^{N-1} T_m}
\right]
\end{equation}
($k<N$), with $\overline T\equiv\sum_{n=0}^{N-1} T_n/N$.  
Due to the
broken symmetry upon profile reflection with respect to the vertical axis passing 
through $N/2$, in our exemplification we
may distinguish two cases for each temperature profile: One in which
the hotter temperatures are applied at the side of the free end in $n=0$
(as in Fig.~\ref{fig_profile}) and one in which
hot temperatures are applied at the side of the fixed mass in $n=N$ 
(perform the transformation $T_n\mapsto T_{N-n}$ to the profiles in Fig.~\ref{fig_profile}). 
Results are respectively depicted in
Fig.~\ref{fig_4cases}(c) and \ref{fig_4cases}(d). In both cases, even the linear
temperature profile does not lead to equipartition. From the plots
one notices that low-$k$ modes store more energy if the free end
is hotter.  This alludes to suggestive implications: The mixed
boundary is the case considered in Ref.~\cite{conti_01}, where
an experiment with a solid bar and a numerical study of an anharmonic
chain displayed behaviors qualitatively consistent with that of
Fig.~\ref{fig_4cases}(c).  Our results thus suggest that the noise at lowest $k$'s
would be lowered by letting the free end to float in a colder
environment. This also points out a conceivable indication for reducing the
measured thermal noise in experiments passible to schematizations analogous to 
those in Fig.~\ref{fig_sketch}.

\begin{figure}
\includegraphics[width=.78\columnwidth]{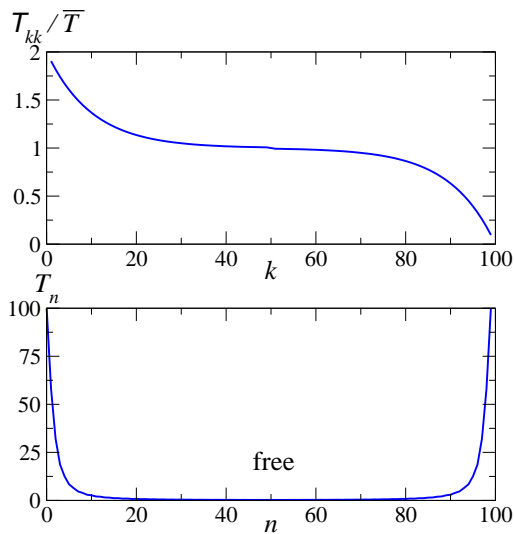}\\
\caption{
  Reconstruction of the temperature profile (lower panel) through
  Eq.~\eqref{eq_rev}, starting from ${\cal T}_{k k}/\overline T$ displayed
  in the upper panel and $\overline T=5.5$. 
}
\label{fig_free_rev}
\end{figure}

\section{Reverse engineering}

The expression ${\cal T}_{k k}(T_n)$ may be inverted, 
thus determining which heat-bath temperature profiles $T_n$
may correspond to a given mode energy repartition. 
For definiteness, let us focus on the case of free boundaries. 
Thanks to simple identities (Appendix~\ref{app:C}),
the inversion of Eq.~\eqref{Tkk} gives
\begin{equation}
\label{eq_rev}
T_n+T_{N-n}=
2\sum_{k=1}^N
\cos\left(\frac{2n+1}{N+1}\,k\pi \right){\cal T}_{k k}
+2\overline T.
\end{equation}
Notice that, given ${\cal T}_{k k}$, the temperature profile $T_n$ is not
uniquely identified. In fact, 
the relation ${\cal T}_{k k}(T_n)$ is many-to-one --
for instance, 
already on the basis of symmetry one can figure out that  
temperature profiles $T_n,\;T'_n$ related by the transformation $T'_n = T_{N-n}$ produce 
the same energy repartition $T_{k k}$.
In the lower panel of 
Fig.~\ref{fig_free_rev} we display a profile reconstruction originated from 
the specific choice for ${\cal T}_{k k}$ reported in the upper one. 
For simplicity, we complemented Eq.~\eqref{eq_rev} with 
the condition $T_n=T_{N-n}$; this means, in particular, $T_0=T_N$.
Our example points out that in principle it is possible to design
heat-bath temperature profiles so that the energy stored in the normal modes of 
the chain is arbitrarily distributed in the range 
$[0,2 \overline T]$, consistently with the condition
$\sum_{k=1}^N{\cal T}_{k k}=N\,\overline T$.

\section{Power spectrum}

To show that $\mathcal{T}_{k k}$ also encodes
the dynamics of fluctuations, we compute
the power spectrum $S(\omega)$ of the chain length $R_N-R_0+N\,l_0$ in
the frequency domain, a quantity typically monitored in
experiments~\cite{conti_01}.  According to the Wiener-Khinchin theorem
\cite{kuboII}, under stationary conditions $S(\omega)$ is given by the
Fourier transform of the chain length's autocorrelation function.
Referring again to free boundary conditions, in terms of normal modes
we have ${R_N-R_0 =\sum_{k=1}^{N}(\Phi_{N k}-\Phi_{0k})X_{k}}$.
Hence (Appendix~\ref{app:D}),  
\begin{align}
\label{eq_pow_spect}
S(\omega)&=
\!
\frac{2\gamma}{N+1}
\sum_{k,k'=1}^N
\frac{
  \;(\Phi_{Nk}-\Phi_{0k})
  \;(\Phi_{Nk'}-\Phi_{0k'})
  \;\mathcal{T}_{k{k'}}
}{
  ({\omega_k^2}-\omega^2-\mathrm{i}\gamma\omega)
  \;({\omega_{k'}^2}-\omega^2+\mathrm{i}\gamma\omega)
}
\\
\label{eq_pow_spect2}
&\simeq
\frac{16\gamma}{N+1}
\sum_{\textrm{odd}\; k}
\frac{
  \cos^2\left(\frac{k}{2(N+1)}\,\pi\right)
           \,{\cal T}_{k k}}
     {(\omega_{k}^2-\omega^2)^2+\gamma^2\,\omega^2}
\end{align}
($\omega\neq0$),
where even modes do not contribute owing to the symmetry of the boundaries. 
Eq.~\eqref{eq_pow_spect2} neglects
the cross-correlations between modes at different $k$.
Such cross-correlation terms are instead responsible for the 
heat flux along the chain, $J_n= l_0\,\kappa\,\langle \dot R_n R_{n-1} \rangle$ \cite{lepri_03}.
In terms of normal modes we have (Appendix~\ref{app:D}) in fact
\begin{align}
J_n= -&
\frac{\mathrm{i}\,l_0\,\kappa}{(2\pi)^2} 
\;\frac{2\gamma}{N+1}
\sum_{k \neq k'} 
\Phi_{n k} \Phi_{n-1, k'} \;\mathcal{T}_{k{k'}}
\times
\nonumber\\ &
\times\int
\text{d} \omega\,
\;\frac{
  \omega
}{
  ({\omega_k^2}-\omega^2-\mathrm{i}\gamma\omega)
  \;({\omega_{k'}^2}-\omega^2+\mathrm{i}\gamma\omega)
}
\end{align}
($0<n<N$).
We have checked that the contribution of terms with $k'\ne k$ in Eq.~\eqref{eq_pow_spect}
can only be appreciated in proximity of the negative peaks of the power spectrum,
away from  the resonances $\omega_k$.

\section{Conclusions}

In summary, our analytic study of energy repartition in a harmonic 
chain in contact with independent heat baths shows that
both long and short wavelength modes may have energies which deviate significantly from 
the level expected if equipartition were to hold. This enhanced or reduced storage of energy 
depends critically on the shape of the temperature profile and on the boundary conditions.
Other dynamical properties, such as the damping or the elastic coupling, are instead totally irrelevant.
Thus, for a generic harmonic chain, information encoded in the temperature profile is mapped into
a sequence of vibrational mode temperatures, which
in turn shape the power spectrum of the chain length.
Investigations about the influence on the above picture
of nonlinearities originating thermo-mechanical couplings, 
is the next important step.

\section*{Acknowledgments}
We would like to thank G. Benettin, S. Lepri, and R. Livi for useful discussions. 
We also thank the Galileo Galilei Institute for Theoretical Physics
for the hospitality and the INFN for partial support during the
completion of this work.

\appendix

\section{Boundary conditions}
\label{app:A}

\subsection{Free boundary conditions}

In the case of free boundary conditions with $N+1$ oscillators the Laplacian matrix is 
\begin{equation}
A 
\equiv
\left(A_{nm}\right)_{n,m=0,1,\ldots,N}
\equiv
\left( \begin{array}{cccccc}
 +1 & -1 &  0 & \cdots & & 0\\
-1 &  +2 & -1 &  \ddots & & \vdots \\
0 & -1 & \ddots & \ddots & & \\
 \vdots& \ddots &  \ddots &  & -1 & 0 \\
  &  &  & -1 & +2 & -1 \\
0  & \cdots  &  & 0 & -1 & +1 
\end{array} \right)\;,
\end{equation}
which is diagonalized by the linear transformation $\Phi^{-1}A\Phi$, with
\begin{equation}
  \Phi_{nk}= 
\left\{ \begin{array}{ll}
1& k=0\\
\sqrt{2} \cos\left( \frac{(2n+1)\;k}{2(N+1)}\;\pi\right) & k\neq0
\end{array} \right.\;,
\end{equation}
\begin{equation}
  \Phi^{-1}=\frac{\Phi^t}{N+1}\;. 
\end{equation}
It is straightforward to see that the definition
\begin{equation}
\mathcal{T}_{kk}\equiv
(N+1)
\sum_{n=0}^N
\Phi^{-1}_{kn}\,\Phi^{-1}_{kn}\,T_n
\end{equation}
leads to
\begin{equation}
  {\cal T}_{kk}=
{\overline T}
\,\left[
1+ \frac{\sum_{n=0}^N T_n
\cos\left(\frac{2n+1}{N+1}\,k\pi \right)}{\sum_{m=0}^N T_m}
\right]
\quad(0<k\leq N)
\end{equation}
and ${\cal T}_{00}=\overline T\equiv\sum_{n=0}^N T_n/(N+1)$.
Notice that $\sum_{n=0}^N \cos\left(\frac{2n+1}{N+1}\,k\pi \right)=0$, 
so that at equilibrium, $T_n=\overline T\;\forall n$, we recover $\mathcal{T}_{kk}=\overline T\;\forall k$.

\subsection{Fixed boundary conditions}

In the case of fixed boundary we can stick to our notations by fixing the
two masses at the border. The number of oscillators becomes $N-1$, and the Laplacian matrix reads
\begin{equation}
A 
\equiv
\left(A_{nm}\right)_{n,m=0,1,\ldots,N}
\equiv
\left( \begin{array}{cccccc}
 0 & 0 &  0 & \cdots & & 0\\
0 &  +2 & -1 &  \ddots & & \vdots \\
0 & -1 & \ddots & \ddots & & \\
 \vdots& \ddots &  \ddots &  & -1 & 0 \\
  &  &  & -1 & +2 & 0 \\
0  & \cdots  &  & 0 & 0 & 0 
\end{array} \right)\;,
\end{equation}
which now is diagonalized by 
\begin{equation}
  \Phi_{nk}= 
\left\{ \begin{array}{ll}
\sqrt{N}& (n,k)=(0,0)\;\textrm{or}\\
  & (n,k)=(N,N) \\
0 & (0<n\leq N,k=0)\;\textrm{or}\\
  & (0<n\leq N,k=N) \\
\sqrt{2} \sin\left( \dfrac{n\;k}{N}\;\pi\right) & \textrm{otherwise}
\end{array} \right.\;,
\end{equation}
\begin{equation}
  \Phi^{-1}=\frac{\Phi^t}{N}\;. 
\end{equation}
Also in this case it is straightforward to show that
\begin{align}
  {\cal T}_{kk}=
\frac{
(N-1)\,
{\overline T}
}{
N}
\,\left[
1- \frac{\sum_{n=1}^{N-1} T_n
\cos\left(\frac{2n\,k}{N}\,\pi \right)}{\sum_{m=1}^{N-1} T_m}
\right] & \\
\qquad(0<k<N), & \nonumber
\end{align}
with $\overline T\equiv\sum_{n=1}^{N-1} T_n/(N-1)$.
We have
$\sum_{n=1}^{N-1}\cos\left(\frac{2n\,k}{N}\,\pi \right)=-1$, 
so that at equilibrium we again recover $\mathcal{T}_{kk}=\overline T\;\forall k$.

\subsection{Mixed boundary conditions}

In the case of mixed boundary we fix only the mass at $n=N$. 
The number of oscillators becomes $N$ and the Laplacian matrix is
\begin{equation}
A 
\equiv
\left(A_{nm}\right)_{n,m=0,1,\ldots,N}
\equiv
\left( \begin{array}{cccccc}
 +1 & -1 &  0 & \cdots & & 0\\
-1 &  +2 & -1 &  \ddots & & \vdots \\
0 & -1 & \ddots & \ddots & & \\
 \vdots& \ddots &  \ddots &  & -1 & 0 \\
  &  &  & -1 & +2 & 0 \\
0  & \cdots  &  & 0 & 0 & 0 
\end{array} \right)\;,
\end{equation}
which is diagonalized by the linear transformation $\Phi^{-1}A\Phi$ with
\begin{equation}
  \Phi_{nk}= 
\left\{ \begin{array}{l}
\sqrt{\frac{2N+1}{2}}\qquad (n,k)=(N,N) \\
0 \qquad\quad\qquad (n=N,0\leq k< N)\;\textrm{or}\\
\ \qquad\quad\qquad (0\leq n<N,k=N) \\
\sqrt{2} \cos\left( \dfrac{(2n+1)\;(2k+1)}{2(2N+1)}\;\pi\right)\quad  \textrm{otherwise}
\end{array} \right.\;,
\end{equation}
\begin{equation}
  \Phi^{-1}=\frac{2\,\Phi^t}{2N+1}\;. 
\end{equation}
As for the previous cases, it is easy to prove that
\begin{align}
  {\cal T}_{kk}=
\frac{
2N\,{\overline T}
}{
2N+1
}
\,\left[
1+ \frac{\sum_{n=0}^{N-1} T_n
\cos\left(\frac{(2n+1)\,(2k+1)}{2N+1}\,\pi \right)}{\sum_{m=0}^{N-1} T_m}
\right] & \\
(0\leq k<N),&\nonumber
\end{align}
with $\overline T\equiv\sum_{n=0}^{N-1} T_n/N$.
In this case
$\sum_{n=0}^{N-1}\cos\left(\frac{(2n+1)\,(2k+1)}{2N+1}\,\pi \right)=1/2$, 
and again one recovers $\mathcal{T}_{kk}=\overline T\;\forall k$ at equilibrium.

\section{Energy repartition among the modes}
\label{app:B}

Equation 
\begin{equation}\label{LangevinX_app}
\ddot X_k= -\gamma\,\dot X_k -\omega_k^2\, X_k + F_k
\end{equation}
is a first-order linear differential
equation in the vector $\mathbf Y_k\equiv(X_k,\dot X_k)$.
Its stationary solution is formally given by
\begin{equation}\label{harmonic}
\mathbf Y_k(t)= \int_{-\infty}^{t}\;dt' \exp[(t-t')\mathbf \Lambda_k] \cdot \mathbf F_k(t'),
\end{equation}
with the definitions
\begin{align}
& \mathbf \Lambda_k= 
\begin{pmatrix}
    0 & 1 \\
    -\omega_k^2 & -\gamma \\
  \end{pmatrix},
&& \mathbf F_k=
   \begin{pmatrix}
    0\\
    F_k\\
  \end{pmatrix}.
\end{align}
The matrix exponential in Eq.~\eqref{harmonic} is computed 
by diagonalizing $\mathbf\Lambda_k$. 
Its eigenvalues $\lambda^{\scriptscriptstyle{1,2}}_k$ are the two 
solutions of the characteristic equation for the unforced harmonic
oscillator, namely
\begin{align}
&\lambda^\alpha_k =
\frac{1}{2} \left( 
-\gamma +(-1)^{\alpha-1} \sqrt{\gamma^2 -4\,\omega_k^2}
\right), 
&&\alpha=1,2\,.
\end{align}
Therefore, from the solutions
\begin{eqnarray}
X_k(t)&=& \sum_{\alpha=1,2}
\int_{-\infty}^{t} \,dt'\,A_k^\alpha \exp(-\lambda_k^\alpha(t-t')) \,F_k(t'),\qquad
\\
\dot X_k(t)&=& \sum_{\alpha=1,2}\int_{-\infty}^{t}\,dt'\, B_k^\alpha \exp(-\lambda_k^\alpha(t-t')) \,F_k(t'),\qquad 
\end{eqnarray}
with 
\begin{eqnarray}
A^1_k&=&\frac{1}{\lambda_k^2-\lambda_k^1}=-A_k^2, 
\\ 
B^1_k&=&-\frac{\lambda_k^1}{\lambda_k^2} B^2_k= \lambda_k^1A_k^2,
\end{eqnarray}
we can evaluate the stationary equal-time correlations 
\begin{align}
&\langle X_k X_{k'}\rangle= 2 \zeta \mathcal T_{kk'}\sum_{\alpha,\beta=1,2} \frac{A_k^\alpha A_{k'}^\beta}{\lambda_k^\alpha+\lambda_{k'}^\beta}, \label{Pos}\\
&\langle\dot X_k \dot X_{k'}\rangle= 2 \zeta \mathcal T_{kk'}\sum_{\alpha,\beta=1,2} \frac{B_k^\alpha B_{k'}^\beta}{\lambda_k^\alpha+\lambda_{k'}^\beta} \label{Vel}.
\end{align}
For the average
kinetic and potential energy per mode, 
\begin{eqnarray}
K_k&\equiv&(N+1)\,\langle \dot{X}_k^2\rangle\,(1-\delta_{k0}/2) ,\\
V_k&\equiv&(N+1)\,\omega_k^2\,\langle X_k^2\rangle, 
\end{eqnarray}
we thus obtain the basic result 
\begin{equation}
K_k^X=V_k^X={\cal T}_{kk}/2\quad(k\neq0).
\end{equation}

\section{Reconstructing the temperature profile}
\label{app:C}

Expressing the cosine in complex notation it is easy to prove the following identities:
\begin{align}
&\sum_{k=1}^N
\cos\left(\frac{(2m+1)\,k}{N+1}\,\pi \right)
\;\cos\left(\frac{(2n+1)\,k}{N+1}\,\pi \right)
=\nonumber\\
&\qquad= \frac{N+1}{2}
\left(\delta_{m\,n}+\delta_{m\,N-m}\right)
-1,
\end{align}
\begin{equation}
\sum_{k=1}^N
\cos\left(\frac{(2m+1)\,k}{N+1}\,\pi \right)
=0.
\end{equation}
Hence, from Eq.~(6) we obtain
\begin{align}
&\sum_{k=1}^N
\cos\left(\frac{(2m+1)\,k}{N+1}\,\pi \right)
\;{\cal T}_{kk}
=\\
&\qquad = \sum_{k=1}^N
\cos\left(\frac{(2m+1)\,k}{N+1}\,\pi \right)
{\overline T}\nonumber\\
&\qquad\qquad \times
\,\left[
1+ \frac{\sum_{n=0}^N T_n
\cos\left(\frac{2n+1}{N+1}\,k\pi \right)}{\sum_{m=0}^N T_m}
\right],
\end{align}
or
\begin{equation}
T_n+T_{N-n}=
2\sum_{k=1}^N
\cos\left(\frac{2n+1}{N+1}\,k\pi \right){\cal T}_{kk}
+2\overline T.
\end{equation}

\section{Spectral density}
\label{app:D}

According to the Wiener-Khinchin theorem
\cite{kuboII}, under stationary conditions
the spectral density $S(\omega)$ of the chain length 
$R_N-R_0+N\,l_0$
is given by
\begin{widetext}
\begin{equation}
S(\omega)=\int\diff \tau\;\mathrm{e}^{\mathrm{i}\omega t}
\left\langle
\;\left[R_N(t_0)- R_0(t_0)\right]
\;\left[R_N(t_0+\tau)- R_0(t_0+\tau)\right]
\right\rangle
+2\pi\, N^2\,l_0^2\;\delta(\omega)
\end{equation}
We have
\begin{equation}
R_N(t)- R_0(t)= \sum_{k=1}^N (\Phi_{Nk}-\Phi_{0k})\;X_k(t),
\end{equation}
so that 
\begin{equation}
S(\omega)=
\sum_{k,k'=1}^N
\;(\Phi_{Nk}-\Phi_{0k})
\;(\Phi_{Nk'}-\Phi_{0k'})
\int\diff \tau\;\mathrm{e}^{\mathrm{i}\omega t}
\left\langle
\;X_k(t_0)
\;X_{k'}(t_0+\tau)
\right\rangle
+2\pi\, N^2\,l_0^2\;\delta(\omega)
\end{equation}
\end{widetext}
We indicate the
Fourier transform of a generic function $h(t)$ as 
$\displaystyle{\widehat  h(\omega)\equiv \int \diff t\,\mathrm{e}^{\text{i} \omega t} \,h(t)}$, and denote
its complex conjugate as $\widehat h^*(\omega)$. 
The Fourier transform of Eq.~\eqref{LangevinX} gives
\begin{equation}
- \omega^2 \widehat X_k(\omega)= \mathrm{i} \omega \gamma \widehat
X_k(\omega) -{\omega_k^2} \widehat X_k(\omega) + \widehat
\eta_k(\omega).
\end{equation}
Solving for $\widehat X_k(\omega)$ and using 
\begin{equation}
\langle \eta_k(t)\,\eta_{k'}(t')\rangle=2\gamma\,\mathcal{T}_{k k'}\,\delta(t-t')/(N+1)
\end{equation}
we obtain, for $\omega\neq0$,
\begin{equation}\label{spectralX}
S(\omega)=
\frac{2\gamma}{N+1}
\sum_{k,k'=1}^N
\frac{
  \;(\Phi_{Nk}-\Phi_{0k})
  \;(\Phi_{Nk'}-\Phi_{0k'})
  \;\mathcal{T}_{k{k'}}
}{
  ({\omega_k^2}-\omega^2-\mathrm{i}\gamma\omega)
  \;({\omega_{k'}^2}-\omega^2+\mathrm{i}\gamma\omega)
}.
\end{equation}

The local heat flux $J_n$ along the chain \cite{lepri_03} is given by
\begin{equation}
J_n=l_0\,\kappa \;\mean{\dot R_n R_{n-1}}\qquad(0<n<N).
\end{equation}
In terms of normal modes the local heat flux becomes
\begin{equation}
J_n=l_0\,\kappa\;\sum_{k\neq k'} \Phi_{nk}\,\Phi_{n-1 k'} \;\mean{\dot X_k X_{k'}}.
\end{equation}
Indeed, stationarity implies $\mean{\dot X_k X_{k}}=0$ for equal-time averages. 
We then have 
\begin{widetext}
\begin{align}
\mean{\dot X_k X_{k'}}&
= \frac{1}{(2 \pi)^2}\int \diff \omega \int
\diff \omega' (- \text{i} \omega) 
\;\mathrm{e}^{-\text{i} \omega t} 
\;\mathrm{e}^{\text{i}\omega' t} 
\mean{ \widehat X_k(\omega) \;\widehat X^*_{k'}(\omega')}
\\ &
= \frac{1}{(2 \pi)^2}\int \diff \omega \int
\diff \omega' (- \text{i} \omega) 
\;\mathrm{e}^{-\text{i} \omega t} 
\;\mathrm{e}^{\text{i}\omega' t} 
\;\frac{2\gamma}{N+1}
\;\frac{
  \;\mathcal{T}_{k{k'}}
}{
  ({\omega_k^2}-\omega^2-\mathrm{i}\gamma\omega)
  \;({\omega_{k'}^2}-\omega'^2+\mathrm{i}\gamma\omega')
}
\;\delta(\omega-\omega')
\\ &
= - \frac{\text{i}}{(2 \pi)^2} 
\;\frac{2\gamma\;\mathcal{T}_{k{k'}}}{N+1}
\int \diff \omega 
\;\frac{
  \omega
}{
  ({\omega_k^2}-\omega^2-\mathrm{i}\gamma\omega)
  \;({\omega_{k'}^2}-\omega^2+\mathrm{i}\gamma\omega)
}.
\end{align}
Putting things together we obtain
\begin{align}
J_n= 
-\frac{\mathrm{i}\,l_0\,\kappa}{(2\pi)^2} 
\;\frac{2\gamma}{N+1}
\sum_{k \neq k'} 
\Phi_{n k} \Phi_{n-1, k'} \;\mathcal{T}_{k{k'}}
\int
\text{d} \omega\, 
\;\frac{
  \omega
}{
  ({\omega_k^2}-\omega^2-\mathrm{i}\gamma\omega)
  \;({\omega_{k'}^2}-\omega^2+\mathrm{i}\gamma\omega)
}
\qquad(0<n<N).
\end{align}
\end{widetext}


\begin{thebibliography}{99}

\bibitem{deZarate_2006} 
  J. M. Ortiz de Z\'arate and J. V. Sengers,
  {\it Hydrodynamic Fluctuations in Fluids and Fluid Mixtures}, Elsevier,
  New York (2006).

\bibitem{Schmitz.1988}
R. Schmitz, {\it Fluctuations in Nonequilibrium Fluids}, Physics Reports 171, 1 (1988).

\bibitem{Sengers_PRA1992}
  P. N. Segr\`e, R.W. Gammon, J. V. Sengers, and  B. M. Law,
  {\it Rayleigh scattering in a liquid far from thermal equilibrium},
  Phys. Rev. A 45, 714
  (1992).

\bibitem{Sengers_PRL1998}
  W. B. Li, K. J. Zhang, J. V. Sengers, R. W. Gammon, and  J. M. Ortiz de Z\'arate,
  {\it Concentration Fluctuations in a Polymer Solution under a Temperature Gradient},
  Phys. Rev. Lett. {\bf 81}, 5580
  (1998).

\bibitem{Takacs_PRL2011}
  C. J. Takacs, A. Vailati, R. Cerbino, S. Mazzoni, M. Giglio, and D. S. Cannell,
  {\it Thermal Fluctuations in a Layer of Liquid ${\mathrm{CS}}_{2}$
    Subjected to Temperature Gradients with and without the Influence
    of Gravity}, 
  Phys. Rev. Lett. {\bf 106}, 244502
  (2011). 

\bibitem{Harris.2008}
 A.~C.~Bleszynski-Jayich, W.~E.~Shanks, and J.~G.~E.~Harris, 
{\it Noise thermometry and electron thermometry of a sample-on-cantilever 
system below 1Kelvin}, 
Applied Physics Letters {\bf 92}, 013123 (2008).

\bibitem{conti}
  L. Conti, M. Bonaldi, L. Rondoni,
  {\it RareNoise: non-equilibrium effects in detectors of gravitational waves }, 
  Cl. Quant. Grav. {\bf 24}, 084032 (2010).

\bibitem{conti_01}
  L. Conti, P. De Gregorio, G. Karapetyan, C. Lazzaro, M. Pegoraro,
  M. Bonaldi, L. Rondoni,
  {\it Effects of breaking vibrational energy equipartition on
  measurements of temperature in macroscopic oscillators subject to
  heat flux}, 
  J. Stat. Mech. P12003 (2013).

\bibitem{bellon}
F. Aguilar Sandoval, M. Geitner, E. Bertin, L. Bellon, 
{\it Resonance frequency shift of strongly heated micro-cantilevers}
J. Appl. Phys. {\bf 117}, 234503 (2015)

\bibitem{monotonic_dispersion}
  In order to
  identify low frequency fluctuations with low wavenumber fluctuations,
  one has to assume a monotonic dispersion relation, which is the case
  for acoustic phonons.

\bibitem{lepri_97} 
  S. Lepri, R. Livi, and A. Politi
  {\it Heat Conduction in Chains of Nonlinear Oscillators},
  Phys. Rev. Lett. {\bf 78}, 1896 (1997)


\bibitem{lepri_03} 
  S. Lepri, R. Livi, and A. Politi
  {\it Thermal conduction in classical low-dimensional lattices}, 
  Phys. Rep. {\bf 377}, 1
  (2003). 

\bibitem{dhar_08}
A. Dhar,
{\it Heat transport in low-dimensional systems}
Adv. in Phys. {\bf 57}, 457 (2008).


\bibitem{bolsterli_01}
M. Bolsterli, M. Rich, W. M. Visscher, 
{\it Simulation of Nonharmonic Interactions in a Cristal by Self-Consistent Reservoirs},
Phys. Rev. A {\bf 1}, 1086 (1970).

\bibitem{bonetto_01}
F. Bonetto, J. L. Lebowitz, J. Lukkarinen, 
{\it Fourier's Law for a Harmonic Crystal with Self-Consistent Stochastic Reservoirs},
J. Stat. Phys. {\bf 116}, 783 (2004).

\bibitem{dhar_06}
D. Dhar and D. Roy,
{\it Heat Transport in Harmonic Lattices},
J. Stat. Phys. {\bf 125}, 801 (2006).

\bibitem{falcao_06}
E. Pereira and R. Falcao,
{\it Normal Heat Conduction in a Chain with a Weak Interparticle Anharmonic Potential},
Phys. Rev. Lett. {\bf 96}, 100601 (2006).

\bibitem{fogedby_14}
H.C. Fogedby and A. Imparato,
{\it Heat fluctuations and fluctuation theorems in the case of multiple
reservoirs},
J. Stat. Mech. P11011 (2014).

\bibitem{imparato_15}
F. Nicacio, A. Ferraro, A. Imparato, M. Paternostro, and F. L. Semi\~ao,
{\it Thermal transport in out-of-equilibrium quantum harmonic chains},
Phys. Rev. E {\bf 91}, 042116 (2015).

\bibitem{Kagra}
K. Somiya,
{\it Detector configuration of KAGRA - the Japanese cryogenic gravitational-wave detector}
Class. Quant. Grav. {\bf 29} 124007 (2012).

\bibitem{continuous} 
  This can be viewed as a simplified description of a
  continuous background temperature field as long as the
  characteristic length scale of the temperature variations are large
  with respect to the typical fluctuations of the oscillator
  positions, i.e. $T/\nabla T|_{n} \gg \sqrt{\langle R_n^2 \rangle}$.

\bibitem{kundu_10}
A. Kundu, A. Chaudhuri, D. Roy, A. Dhar, J. L. Lebowitz, and H. Spohn, 
{\it Heat conduction and phonon localization in disordered harmonic crystals},
Europhys. Lett. {\bf 90}, 40001 (2010).

\bibitem{kuboII}
  See, e.g., 
  M. Toda, N. Saito, and R. Kubo,
  {\it Statistical Physics II: Nonequilibrium Statistical Mechanics},
  (Springer Series in Solid-State Sciences 1991).



\end{thebibliography}
\end{document}